\documentclass[a4paper]{article}

\usepackage{INTERSPEECH2021}
\usepackage{comment}
\usepackage{multirow}

\usepackage{amsmath,graphicx}
\usepackage{siunitx}
\usepackage[acronym]{glossaries}

\usepackage{pgfplots}
\usepackage{pgfplotstable}

\usepackage{balance}

\usepackage{tikz}
\usetikzlibrary{calc}
\usetikzlibrary{positioning}
\usetikzlibrary{chains}
\usetikzlibrary{fit}
\usetikzlibrary{arrows}
\usetikzlibrary{decorations.pathreplacing}
\usetikzlibrary{calligraphy}
\usetikzlibrary{arrows.meta}
\usetikzlibrary{backgrounds}
\usetikzlibrary{tikzmark}
\usetikzlibrary{shapes.multipart}
\usetikzlibrary{matrix}
\pgfplotsset{compat=1.3}

\usepgfplotslibrary{groupplots}
\tikzset{
>=stealth,
element/.style={
draw=black!100,
thick,
align=center,
inner sep=0pt,
},
narrow block/.style={
element,
rectangle,
text width=4.0em,
minimum width=4.0em,
minimum height=2.5em,
},
block/.style={
element,
rectangle,
text width=7.0em,
minimum width=7.0em,
minimum height=2.5em,
},
branch/.style={
element,
circle,
fill=black,
minimum size=0.2em,
},
apply/.style={
element,
circle,
minimum size=1em,
label=center:{$\times$}
},
add/.style={
element,
circle,
minimum size=1em,
label=center:{$+$}
},
arrow/.style={->, shorten >=0.1em},
reverse arrow/.style={<-, shorten <=0.1em},
bidirectional arrow/.style={<->, shorten <=0.1em, shorten >=0.1em},
}

\usepackage[capitalise, noabbrev]{cleveref}

\newacronym{AM}{AM}{acoustic model}
\newacronym{ASR}{ASR}{automatic speech recognition}
\newacronym{DFT}{DFT}{discrete Fourier transform}
\newacronym{KLT}{KLT}{Karhunen-Loève Transform}
\newacronym{LFR}{LFR}{low frame rate}
\newacronym{nWER}{nWER}{normalized word error rate}
\newacronym{WER}{WER}{word error rate}
\newacronym{WERR}{WERR}{word error rate reduction}
\newacronym{rWERR}{rWERR}{relative word error rate reduction}
\newacronym{MVDR}{MVDR}{Minimum Variance Distortionless Response}
\newacronym{RNN-T}{RNN-T}{recurrent neural network transducer}
\newacronym{STFT}{STFT}{short-time Fourier transformation}
\newacronym{VAD}{VAD}{voice activity detection}
\newacronym{VoIP}{VoIP}{voice over IP}
\newacronym{tf-bin}{tf-bin}{time-frequency bin}
\newacronym{SNR}{SNR}{signal-to-noise ratio}
\newacronym{IPD}{IPD}{inter-aural phase difference}
\newacronym{LSTM}{LSTM}{long short-term memory}
\newacronym{rAELR}{rAELR}{relative average emission latency reduction}

\title{Contextual-Utterance Training for Automatic Speech Recognition}
\name{Alejandro Gomez-Alanis$^1$, Lukas Drude$^1$, Andreas Schwarz$^1$, Rupak Vignesh Swaminathan$^2$, Simon Wiesler$^1$}
\address{$^1$Amazon Alexa, Aachen, Germany; $^2$Amazon Alexa, Pittsburgh, USA}
\email{llnis@amazon.com, drude@amazon.com, asw@amazon.com, swarupak@amazon.com, wiesler@amazon.com}

\begin{document}

\maketitle
\begin{abstract}

Recent studies of streaming \gls{ASR} \gls{RNN-T}-based systems have fed the encoder with past contextual information in order to improve its \gls{WER} performance. In this paper, we first propose a contextual-utterance training technique which makes use of the previous and future contextual utterances in order to do an implicit adaptation to the speaker, topic and acoustic environment. Also, we propose a dual-mode contextual-utterance training technique for streaming \gls{ASR} systems. This proposed approach allows to make a better use of the available acoustic context in streaming models by distilling ``in-place'' the knowledge of a teacher (non-streaming mode), which is able to see both past and future contextual utterances, to the student (streaming mode) which can only see the current and past contextual utterances. The experimental results show that a state-of-the-art conformer-transducer system trained with the proposed techniques outperforms the same system trained with the classical \gls{RNN-T} loss. Specifically, the proposed technique is able to reduce both the \gls{WER} and the average last token emission latency by more than \SI{6}{\percent} and \SI{40}{ms} relative, respectively.
\end{abstract}

\section{Introduction}
\label{sec:introduction}

Voice assistants like Amazon Alexa use streaming \gls{ASR} for low-latency recognition of user requests. Streaming \gls{ASR} systems continuously process audio input without requiring ``offline'' processing of full utterances. An example of such a system is the \gls{RNN-T} \cite{rnnt}, and its improved variants where the encoder is based on transformers \cite{transformer_transducer} or conformers \cite{conformer_transducer}.

Modern streaming and non-streaming \gls{ASR} models share most of the neural network architectures since the acoustic, pronunciation, and language models of conventional \gls{ASR} systems have evolved into a single end-to-end neural network such as the \gls{RNN-T} \cite{rnnt,Jain2020ContextualRF}. In this scenario, the dual-mode training \cite{dual_mode} framework has been recently proposed to unify streaming and non-streaming \gls{ASR} networks with shared weights. Sharing network weights while doing an in-place distillation \cite{distillation,swaminathan2021codert,nagaraja2021collaborative} has shown to be effective for streaming \gls{RNN-T} systems in terms of both \gls{WER} and token emission latency.


Most \gls{ASR} systems are designed to recognize independent utterances, despite the fact that contextual information over multiple utterances, such as information on the speaker, topic or acoustic environment, is known to be useful for \gls{ASR} \cite{context1, context2}. There are several approaches to incorporate contextual information in end-to-end \gls{ASR}, such as i-vector approaches that utilize speaker context \cite{audhkhasi2017direct, fan2020speakeraware}. In contrast to such approaches, where a model adapts explicitly to the speaker, more recent works \cite{schwarz2021improving, hori2020} present the entire available audio as context to the \gls{RNN-T} encoder in order to implicitly learn to make use of the available context for acoustic environment and speaker adaptation. 
Moreover, a new method based on triggered attention has been proposed in \cite{hori2021advanced} so that a streaming transformer transducer can make use of contextual information.
However, this technique can only make use of the past contextual information for training the streaming system.

\begin{center}
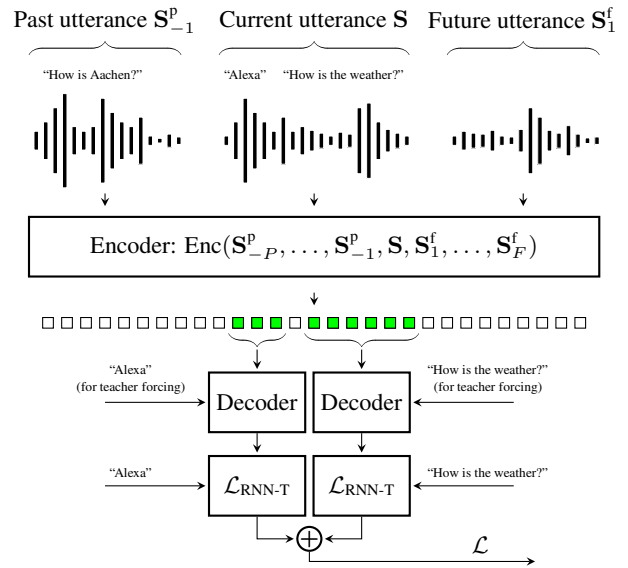
\begin{figure}[t]
\begin{tikzpicture}[on grid, node distance=3.5em and 6.0em]
\coordinate (past left) at (0, 0);
\foreach \x [count=\index] in {1, 2, 3.5, 5, 1.5, 1, 1.5, 4.5, 3, 2, 1.5, 2.5, 0.5, 0.25, 0.7, 0.4}
{
\coordinate (past right) at ($(past left) + (\index/8, 0)$);
\draw[draw=white, fill=black] ($(past right) + (-0.06, -\x/8)$) rectangle ++(0.06,\x/4);
}

\draw [decorate,decoration={calligraphic brace,amplitude=5pt,raise=8ex}]
(past left) -- (past right) node[midway, yshift=45]{\small Past utterance $\mathbf{S}^{\text{p}}_{-1}\strut$} node[midway, xshift=-0.5em, yshift=25]{\tiny ``How is Aachen?''};

\coordinate (current left) at ($(past right) + (1/8+3/8, 0)$);
\foreach \x [count=\index] in {0.5, 2, 4.5, 3, 2, 1, 2.5, 1, 1.5, 1.1, 0.8, 0.5, 0.9, 0.7, 3.5, 4, 2, 1.2, 0.8, 0.5}
{
	\coordinate (current right) at ($(current left) + (\index/8, 0)$);
	\draw[draw=white, fill=black] ($(current right) + (-0.06, -\x/8)$) rectangle ++(0.06,\x/4);
}

\draw [decorate,decoration={calligraphic brace,amplitude=5pt,raise=8ex}]
(current left) -- (current right) node[midway, yshift=45]{\small Current utterance $\mathbf{S}\strut$} node[midway, yshift=25]{\tiny ``Alexa'' \quad ``How is the weather?''};

\coordinate (future left) at ($(current right) + (1/8+3/8, 0)$);
\foreach \x [count=\index] in {0.5, 1, 0.8, 0.7, 1.1, 0.5, 0.4, 1.2, 3.5, 2.5, 1.1, 0.8, 1.6, 0.7, 0.3, 0.4}
{
	\coordinate (future right) at ($(future left) + (\index/8, 0)$);
	\draw[draw=white, fill=black] ($(future right) + (-0.06, -\x/8)$) rectangle ++(0.06,\x/4);
}

\draw [decorate,decoration={calligraphic brace,amplitude=5pt,raise=8ex}]
(future left) -- (future right) node[midway, yshift=45]{\small Future utterance $\mathbf{S}^{\text{f}}_{1}\strut$};

\draw [arrow] ($(past left)!1/2!(past right) + (0, -0.7)$) -- ++(0, -0.2);
\draw [arrow] ($(current left)!1/2!(current right) + (0, -0.7)$) -- ++(0, -0.2);
\draw [arrow] ($(future left)!1/2!(future right) + (0, -0.7)$) -- ++(0, -0.2);

\node (encoder) [block, fit=(past left)(future right), yshift=-40] {};
\node (test) at (encoder.center){Encoder: $\text{Enc}(
	\mathbf{S}^{\text{p}}_{-P}, \dots, \mathbf{S}^{\text{p}}_{-1}, \mathbf{S}, \mathbf{S}^{\text{f}}_{1}, \dots, \mathbf{S}^{\text{f}}_{F}
	)$};

\draw [arrow] ($(encoder.south) + (0, -0.2)$) -- ++(0, -0.2);

\foreach \index in {-14,...,14}
{
\node [draw=black, inner sep=2] (temp) at ($(encoder.south) + (\index/4, -0.6)$) {};
}
\foreach \index in {-4,...,-2}
{
	\node [draw=black, fill=green, inner sep=2] (temp) at ($(encoder.south) + (\index/4, -0.6)$) {};
}
\foreach \index in {0,...,5}
{
\node [draw=black, fill=green, inner sep=2] (temp) at ($(encoder.south) + (\index/4, -0.6)$) {};
}

\draw [decorate,decoration={calligraphic brace,mirror,amplitude=5pt,raise=1ex}] ($(encoder.south) + (-4.5/4, -0.6)$) -- ($(encoder.south) + (-1.5/4, -0.6)$) node (decoder 1) [narrow block, midway, yshift=-30] {Decoder};

\draw [decorate,decoration={calligraphic brace,mirror,amplitude=5pt,raise=1ex}] ($(encoder.south) + (-0.5/4, -0.6)$) -- ($(encoder.south) + (5.5/4, -0.6)$) node (decoder 2) [narrow block, midway, yshift=-30] {Decoder};

\draw [reverse arrow] (decoder 1) -- +(-2, 0) node [above, near end] {\tiny \shortstack{``Alexa''\\ (for teacher forcing)}};
\draw [reverse arrow] (decoder 2) -- +(2, 0) node [above, near end] {\tiny \shortstack{``How is the weather?''\\ (for teacher forcing)}};

\draw [reverse arrow] (decoder 1) -- +(0, 0.7);
\draw [reverse arrow] (decoder 2) -- +(0, 0.7);

\node (loss 1) [narrow block, below=of decoder 1] {$\mathcal{L}_{\text{RNN-T}}$};
\node (loss 2) [narrow block, below=of decoder 2] {$\mathcal{L}_{\text{RNN-T}}$};

\draw [reverse arrow] (loss 1) -- +(-2, 0) node [above, near end] {\tiny ``Alexa''};
\draw [reverse arrow] (loss 2) -- +(2, 0) node [above, near end] {\tiny ``How is the weather?''};

\draw [arrow] (decoder 1) -- (loss 1);
\draw [arrow] (decoder 2) -- (loss 2);

\node (add) [add, yshift=-20, label=center:{$+$}] at ($(loss 1)!1/2!(loss 2)$) {};
\draw [arrow] (loss 1) |- (add);
\draw [arrow] (loss 2) |- (add);
\draw [arrow] (add) -- ++(0, -0.3) -- ++(3, 0) node [above, near end] {$\mathcal L$};
\end{tikzpicture}
\vspace{-4mm}
\caption{
Block diagram of the proposed contextual-utterance \gls{RNN-T} training.
The \gls{RNN-T} encoder receives all utterances in order to model relations between them and yields the encoder representation.
For each labelled segment of the utterance of interest, the corresponding hidden representations (green) are extracted for the \gls{RNN-T} loss computation.
The decoder only accesses hidden representations of labeled segments within the current utterance.
The overall loss is the sum of losses over all the segments belonging to the current utterance.
}
\label{fig:contextual_stream_training}
\end{figure}
\vspace{-10mm}
\end{center}

In this paper we propose two techniques for training \gls{RNN-T}-based \gls{ASR} systems: (i) contextual-utterance training (depicted in \cref{fig:contextual_stream_training}) which makes use of the available acoustic context utterances for an implicit speaker and acoustic environment adaptation; and (ii) dual-mode contextual-utterance training which shares the weights between streaming and non-streaming modes, trains both streaming and non-streaming modes jointly, and employs an in-place distillation loss in order to distill knowledge from the teacher (non-streaming) to the student (streaming) mode. The core idea of the proposed dual-mode contextual-utterance training is that a streaming system can make better use of acoustic context when distilling the teacher's knowledge about past and future context during training.

The rest of this paper is organized as follows. First, we provide a short review of the classic \gls{RNN-T} \gls{ASR} training in \cref{sec:overview}. Then, we describe our proposed training methods in \cref{sec:method}. Then, in \cref{sec:experimental_setup}, we outline the system details, speech datasets, training configuration and performance metrics. \cref{sec:results} discusses the experimental results. Finally, we summarize the conclusions derived from this research in \cref{sec:conclusions}.

\section{Overview of RNN-T training}
\label{sec:overview}

We employ the RNN-T model architecture to validate the proposed approach due to its recent popularity in ASR systems \cite{rnnt_google_2022, microsoft_rnnt_2019}. The RNN-T model defines the conditional probability distribution $P(\mathbf{y} | \mathbf{x})$ of an output label sequence $\mathbf{y} = [y_1, ..., y_U]$ of length $U$ given a sequence of $T$ feature vectors $\mathbf{x} = [x_1, ..., x_T]$. The classic RNN-T model architecture consists of three distinct modules: an encoder, a prediction network or decoder, and a joint network. The encoder maps sequentially processed feature vectors $[x_1, ..., x_T]$ to high-level acoustic representations, similar to the acoustic model in the hybrid ASR approach: 

\begin{align}
\mathbf{h}  = \text{Enc}(\mathbf{x}).
\end{align}

The prediction network or decoder takes as input a sequence of labels $[y_1 , . . . , y_j]$. The joint network combines the output representations of the encoder and the prediction network and produces activations for each time frame $t$ and label position $j$, which are projected to the output probability distribution $P(\mathbf{y} | \mathbf{x})$ via a softmax layer.

During training, the target label sequence $\mathbf{y}^\star$ is available and used to minimize the negative log-likelihood for a training sample:

\begin{align}
\mathcal{L}_{\text{\gls{RNN-T}}} = - \text{log} P(\mathbf{y}^\star | \mathbf{h}).
\label{eq:loss_rnnt}
\end{align}

In the following, we use $\mathcal{L}_{\text{\gls{RNN-T}}}(\mathbf{h}, \mathbf{y}^\star)$ to denote the computation of the joint network, the prediction network, and the RNN-T loss based on a given encoder output sequence $\mathbf{h}$ and target label sequence $\mathbf{y}^\star$.

\section{Method}
\label{sec:method}

In this section we describe the proposed \gls{ASR} training methods. First, \cref{sec:context_training} outlines the proposed contextual-utterance training method for both non-streaming and streaming \gls{ASR} systems. Then, \cref{sec:dual_mode_context_training} describes the proposed dual-mode contextual-utterance training method for making better use of contextual utterances in streaming systems.

\subsection{Contextual-Utterance Training}
\label{sec:context_training}

We use the term \textit{utterance} or \textit{stream} to refer to the entire audio recording received by the device for one interaction of the user with the voice assistant, which typically includes both an activation keyword (``Alexa'') and the expression of the user intent (``play some music''), and has a typical length of \SI{3}{s} to \SI{15}{s}. Within an utterance, one or multiple speech \textit{segments} may be defined, e.g., by a voice activity detector or by the keyword spotter. We also define the contextual utterances or streams of the utterance of interest as those neighbouring utterances that are either immediate predecessors (past contextual utterances) or follow the current utterance (future contextual utterances). Note that at least one segment of the utterance of interest has to be labeled for training with the proposed method while its contextual utterances can be completely unlabeled.

In order to make use of contextual information, we extend our previous method \cite{schwarz2021improving} where only the entire available feature sequence of the current utterance is forwarded through the encoder followed by the loss computation on individual labeled speech segments. Here, we pass the current utterance as well as its contextual utterances within a predefined window to the encoder which is in charge of modeling the interactions between different utterances and obtaining the encoding representations of the current utterance. Then, the loss is computed for all segments belonging to the current utterance as shown in \cref{fig:contextual_stream_training}.

We use $\mathbf{S}^{\text{p}}_{i}$ and $\mathbf{S}^{\text{f}}_{j}$ to respectively denote the $i$-th past and $j$-th future contextual utterance features with respect to the features of the utterance of interest $\mathbf{S}$. Moreover, we use $P$ and $F$ to denote the number of past and future contextual utterances, respectively. Assuming that we have a non-streaming ASR system which is typically used as a teacher system, its encoder takes as input the features belonging to all utterances $[\mathbf{S}^{\text{p}}_{-P:-1}, \mathbf{S}, \mathbf{S}^{\text{f}}_{1:F}]$ and obtains the encoding sequence for all utterances:
\begin{align}
\mathbf{h}_\text{teacher}  = \text{Enc}(
\mathbf{S}^{\text{p}}_{-P}, \dots, \mathbf{S}^{\text{p}}_{-1}, \mathbf{S}, \mathbf{S}^{\text{f}}_{1}, \dots, \mathbf{S}^{\text{f}}_{F}
).
\end{align}
Now, the part of the encoding sequence which corresponds to a labeled segment is sliced out (compare \cite[Section~3.2]{schwarz2021improving}).
For each segment $u$ of $\mathbf{S}$ with start index $t_{\text{start},u}$ and end index $t_{\text{end},u}$, the corresponding encoding subsequence is selected:
\begin{align}
\label{eq:h_teacher}
\mathbf{h}_{\text{teacher}}^u = \mathbf{h}_\text{teacher}[h_{t_{\text{start},u}}, \dots, h_{t_{\text{end},u}}].
\end{align}
Denoting the segment target label sequence by $\mathbf{y}_u^\star$, the \gls{RNN-T} loss of the teacher model for the segment is
\begin{align}
\mathcal{L}_{\text{teacher}}^u = \mathcal{L}_{\text{\gls{RNN-T}}}(\mathbf{h}_{\text{teacher}}^u , \mathbf{y}_u^\star) = - \text{log} P(\mathbf{y}_u^\star | \mathbf{h}_{\text{teacher}}^u ).
\end{align}

It is worth noticing that this proposed method can be also applied to streaming systems. In order to do it, the encoder can only take as input the past contextual utterances and the current utterance up to the current time step in order to obtain the encoding subsequence in (\ref{eq:h_teacher}).

\subsection{Dual-Mode Contextual-Utterance Training}
\label{sec:dual_mode_context_training}

 The proposed contextual-utterance training method described in the previous section can be used for training either streaming or non-streaming \gls{ASR} models. However, a streaming model is limited to the past contextual utterances and the current utterance up to the current time frame. In order to make better use of contextual utterances in streaming \gls{ASR} models, we aim at combining the proposed contextual-utterance training approach with the dual-mode training method \cite{dual_mode} in order to distill knowledge from the non-streaming mode (teacher) into the streaming mode (student) on the fly within the same model. The main advantage of this idea is the ability to distill the teacher's knowledge (non-streaming mode), which is able to see both past and future contextual utterances, to the streaming mode which can only see the past contextual utterances. 
 
 In dual-mode \gls{ASR} \cite{dual_mode}, the \gls{RNN-T} architecture is trained using two modes: (1) streaming or student mode, and (2) non-streaming or teacher mode. This dual-mode training method unifies and improves streaming \gls{ASR} models by using three main techniques: (1) weights sharing between the streaming and non-streaming modes; (2) joint training of streaming and non-streaming modes for each training iteration; and (3) in-place knowledge distillation \cite{distillation} in order to distill knowledge from the non-streaming mode (teacher) into the streaming mode (student) on the fly within the same model, by encouraging consistency of the predicted token probabilities.

As we said previously, the student encoder can only take as input the past contextual utterances and the current utterance up to the current time step (denoted by the index $t_{\text{current},u}$) to obtain the encoding sequence.
The student and the teacher share all weights:
\begin{align}
\mathbf{h}_\text{student}  = \text{Enc}(
\mathbf{S}^{\text{p}}_{-P}, \dots, \mathbf{S}^{\text{p}}_{-1}, \mathbf{S}
).
\end{align}
Similar to the encoder in non-streaming mode, the encoder in streaming mode extracts the corresponding encoding subsequence $\mathbf{h}_{\text{student}}^u = \mathbf{h}_\text{student}[h_{t_{\text{start},u}}, \dots, h_{t_{\text{current},u}}]$.
Then, the \gls{RNN-T} loss of the streaming mode (student) for the segment $u$ is obtained using that encoding subsequence:
\begin{align}
    \mathcal{L}_{\text{student}}^u = \mathcal{L}_{\text{RNN-T}}(\mathbf{h}_{\text{student}}^u , \mathbf{y}_u^\star) = - \text{log} P(\mathbf{y}_u^\star | \mathbf{h}_{\text{student}}^u ).
    \label{eq:loss_utt}
\end{align}
The overall loss for the utterance of interest $\mathbf{S}$ is the sum of losses over
all the $M$ labeled segments which belong to the utterance of interest.
Thus, the proposed dual-mode contextual-utterance loss for the utterance of interest $\mathbf{S}$ is
\begin{align}
\mathcal{L}_{\text{dual-mode}} = \sum_{u=1}^{M} \mathcal{L}_{\text{teacher}}^u + \mathcal{L}_{\text{student}}^u + \beta \mathcal{L}_{\text{distillation}}^u,
\label{eq:loss_dual}
\end{align}
where $\mathcal{L}_{\text{distillation}}^u$ denotes the in-place knowledge distillation loss for the segment $u$ proposed in \cite{distillation}, and $\beta$ is a hyper-parameter which weighs the importance of the in-place knowledge distillation loss relative to the teacher and student losses.

The in-place knowledge distillation loss~\cite{distillation} is an approximation to the sum of Kullback-Leibler divergences between teacher and student conditional probabilities over the entire \gls{RNN-T} output probability lattice~\cite[Equation~7]{distillation} of one segment.
It here serves as additional coupling (on top of the weight sharing) between the teacher mode and the student mode of the encoder network.

\section{Experimental Setup}
\label{sec:experimental_setup}
In this section we describe the details of the conformer-transducer system as well as the internal datasets and configurations employed in all experiments. 

\subsection{Model Configuration}
\label{sec:model_config}
The \gls{RNN-T}-based system employed in all experiments is a conformer transducer with \SI{73.5}{M} parameters which consists of a conformer encoder, a \gls{LSTM} prediction network and a feed-forward joint network. The conformer encoder has \SI{54.2}{M} parameters arranged in 12 conformer blocks where each one has 8 self-attention heads and where the encoder dimension is 512. The prediction network is a $2 \times 768$ \gls{LSTM} with a total of \SI{15.2}{M} parameters. The joint network is one feed-forward layer with 512 units and tanh activation, followed by a softmax layer with an output vocabulary size of \num{4000} wordpieces.

\subsection{Datasets and Training}
\label{sec:datasets}
All models are trained on an internal dataset of de-identified  recordings from  voice-controlled far-field devices which contains about \SI{50000}{h} of speech. 
While we did not specifically ensure that every utterance has $P$ past and $F$ future context utterances, over \SI{95}{\percent} have at least two past contextual utterances and over \SI{70}{\percent} have at least one future contextual utterance.
In order to evaluate the proposed approach, we define a test dataset for which we ensure that at least $P=2$ past contextual utterances for each utterance of interest are available.
While we did not explicitly ensure that a future contextual-utterance is available in this analysis, most examples still have one future contextual-utterance.

Similar to \cite{hori2021advanced}, we found in our preliminary experiments that it is better to start from a pre-trained model without context training and finetune it by training with contextual streams (warm-start technique). For all trained systems (including the baseline systems without context), we first pre-train the models for a maximum of \SI{550}{k} iterations, and then finetune them by training with $P$ past and $F$ future contextual streams for a maximum of another \SI{1000}{k} iterations.
Note that for a fair comparison, we also train again the baseline systems without context to reach the same number of iterations in total.
Finally, the best checkpoint is chosen based on the \gls{WER} obtained on the validation dataset using the same acoustic context configuration.

We train all systems using the Adam optimizer with a linear warmup period of \SI{5}{k} iterations followed by an exponential decay \cite{kingma2014adam}. We also employ a bucket configuration of $[300, 600, 1000]$ with a respective bucketing batch size of $[4096, 2048, 1024, 256]$ frames~\cite{khomenko2016accelerating}. The acoustic features are 64-dimensional Log-Mel-Frequency features with a frame shift of \SI{10}{ms}, stacked and downsampled by a factor of 3, corresponding to an encoder frame rate of \SI{30}{ms}. We use an adaptive variant of a feature-based augmentation method, SpecAugment \cite{specaugment}, as proposed in \cite{park2019specaugment}.

\subsection{Performance Metrics}
\label{sec:metrics}
The evaluation of the \gls{ASR} systems is done in terms of the \gls{rWERR} (higher is better) as well as in terms of the \gls{rAELR} (higher is better). In all evaluations, we consider the streaming baseline system, i.e., the system trained without using any contextual utterances, as the reference system.

\section{Results}
\label{sec:results}

We first evaluate the baseline streaming system including dual-mode training in \cref{sec:results_dual_mode}. Then, \cref{sec:results_teacher} discusses the results from the proposed contextual-utterance training method for both non-streaming and streaming models. Finally, \cref{sec:results_student} outlines the results from the proposed dual-mode contextual-utterance training technique for streaming systems.

\subsection{Dual-Mode Training Results}
\label{sec:results_dual_mode}

\begin{table}
\centering
\begin{tabular}{
l
c
S[table-format=-1.1,table-text-alignment=center]
S[table-format=-2.1,table-text-alignment=center]
}
\toprule
Model & {$\beta$} & {\!\!\gls{rWERR} (\si{\percent})\!\!} & {\!\!\gls{rAELR} (\si{ms})\!\!} \\
\midrule
Streaming model & 0.0 & 0.0 & 0.0 \\
+ lattice distillation\!\! & {$1\cdot10^{-3}$} & -0.5 & 42.0 \\
+ dual mode & {$5\cdot10^{-3}$} & -1.5 & 83.1 \\
+ dual mode & {$1\cdot10^{-3}$} & 0.8 & 48.3 \\
+ dual mode & {$5\cdot10^{-4}$} & 2.0 & 41.7 \\
+ dual mode & {$1\cdot10^{-4}$} & 1.4 & 19.5 \\
\bottomrule
\end{tabular}
\caption{Tuning results for the lattice distillation loss weight $\beta$ on an independent development set without contextual utterances.}
\label{table:beta}
\end{table}

\cref{table:beta} shows tuning experiments with dual-mode training to determine the distillation loss weight $\beta$ on an independent development set.
Based on that analysis, we have chosen a distillation loss weight of $\beta = 5 \cdot 10^{-4}$. A very high value, e.g., $5 \cdot 10^{-3}$ leads to higher latency gains of the streaming output at the cost of increased \gls{WER}.
A very small $\beta$ leads to almost no coupling between the predictions, thereby reducing the latency improvements of the streaming system.

\begin{table}[b]
\centering
\begin{tabular}{
c
S[table-format=1,table-text-alignment=center]
S[table-format=1,table-text-alignment=center]
S[table-format=2.1,table-text-alignment=center]
}
\toprule
\multirow{2.5}{*}{Model} & \multicolumn{2}{c}{Contextual utterances} & {\multirow{2.5}{*}{\shortstack{\gls{rWERR} \\ (\%)}}} \\
\cmidrule{2-3}
& {Past} & {Future} & \\
\midrule
\multirow{4}{*}{\begin{tabular}[c]{@{}c@{}}\footnotesize Non-streaming model \\[-0.75ex] \footnotesize with contextual utterances\end{tabular}}
& 0 & 0 & 7.6 \\
& 1 & 0 & 11.2 \\
& 2 & 0 & 11.5 \\
& 1 & 1 & 12.1 \\
\midrule
\multirow{4}{*}{\begin{tabular}[c]{@{}c@{}}\footnotesize Streaming model \\[-0.75ex] \footnotesize with contextual utterances\end{tabular}}
& 0 & 0 & 0.0 \\
& 1 & 0 & 2.9 \\
& 2 & 0 & 5.4 \\
& 1 & 1 & 2.9 \\
\midrule
\multirow{4}{*}{\begin{tabular}[c]{@{}c@{}}\footnotesize Dual-mode model \\[-0.75ex] \footnotesize with contextual utterances \\[-0.75ex] \footnotesize evaluated in streaming mode\end{tabular}}
& 0 & 0 & 1.4 \\
& 1 & 0 & 5.2 \\
& 2 & 0 & 6.3 \\
& 1 & 1  & 5.8 \\
\bottomrule
\end{tabular}
\caption{\gls{rWERR} (\%) of the conformer-transducer models trained with either the contextual-utterance or the dual-mode contextual-utterance training technique, with respect to same baseline streaming system trained without contextual utterances. The evaluation of the dual-mode system is done in streaming mode.}
\label{table:dual_mode_context_training_2p_0f}
\end{table}
\begin{table}
\centering
\begin{tabular}{
c
S[table-format=1,table-text-alignment=center]
S[table-format=1,table-text-alignment=center]
S[table-format=2.1,table-text-alignment=center]
}
\toprule
\multirow{2.5}{*}{Model} & \multicolumn{2}{c}{Contextual utterances} & {\multirow{2.5}{*}{\shortstack{\gls{rAELR} \\ (ms)}}} \\
\cmidrule{2-3}
& {Past} & {Future} & \\
\midrule
\multirow{3}{*}{\begin{tabular}[c]{@{}c@{}}\footnotesize Streaming model \\[-0.75ex] \footnotesize with contextual utterances\end{tabular}}
& 0 & 0 & 0.0 \\
& 1 & 0 & -1.5 \\
& 2 & 0 & -2.9 \\
\midrule
\multirow{3}{*}{\begin{tabular}[c]{@{}c@{}}\footnotesize Dual-mode model \\[-0.75ex] \footnotesize with contextual utterances \\[-0.75ex] \footnotesize evaluated in streaming mode\end{tabular}}
& 0 & 0 & 49.3 \\
& 1 & 0 & 44.7 \\
& 2 & 0 & 48.3 \\
\bottomrule
\end{tabular}
\caption{\gls{rAELR} (ms) of the conformer-transducer models trained with either the contextual-utterance or the dual-mode contextual-utterance training technique, with respect to same baseline streaming system trained without contextual utterances. The evaluation of the dual-mode system is done in streaming mode.}
\label{table:dual_mode_context_training_2p_0f_raelr}
\end{table}

\subsection{Contextual-Utterance Training Results}
\label{sec:results_teacher}
The top group in \cref{table:dual_mode_context_training_2p_0f} shows the performance of the teacher (non-streaming) system trained and evaluated with contextual utterances.
The \glspl{rWERR} are calculated with respect to the streaming non-context system (fifth line) to facilitate the comparison with the streaming results.
The systems have been trained using three contextual-utterance configurations: (i) one past contextual utterance, (ii) two past contextual utterances, and (iii) one past and one future contextual utterances.

The proposed contextual-utterance training for the non-streaming model is able to significantly improve the performance of the same system trained without context.
Moreover, \cref{table:dual_mode_context_training_2p_0f} shows that training with two past contextual utterances is better than using only one past contextual utterance.
However, it is consistently better to use one past and one future contextual utterances than only past utterances, achieving up to \SI{12.1}{\percent} of \gls{rWERR}. We hypothesize that these improvements stem from the conformer encoder learning to implicitly adapt to the speaker and/or the acoustic environment based on the additional left and right context seen by the conformer encoder and the higher probability that at least one contextual utterance relates to the utterance of interest.

Moreover, the second group in \cref{table:dual_mode_context_training_2p_0f} shows the performance of the streaming conformer-transducer trained with the proposed contextual-utterance training method. As before, the reference is the streaming system trained without contextual utterances, i.e., it is only trained with the context available in the current utterance as we proposed in our previous work \cite{schwarz2021improving}. We can see that the proposed method achieves up to \SI{5.4}{\percent} of \gls{rWERR} for streaming systems as well. However, using future utterances does not allow to improve its performance since it is a causal system which cannot make use of the future information.

\subsection{Dual-Mode Contextual-Utterance Training Results}
\label{sec:results_student}
The last group in \cref{table:dual_mode_context_training_2p_0f} shows the performance of the streaming conformer-transducer in terms of \gls{rWERR} trained with the proposed dual-mode contextual-utterance training technique. The reference is the same 
streaming system as in \cref{sec:results_teacher} trained without contextual utterances (first line of the second group in \cref{table:dual_mode_context_training_2p_0f}).

First, the dual-mode training technique without using context is able to improve the \gls{WER} by \SI{1.4}{\percent} relative.
However, the proposed dual-mode contextual-utterance training technique is able to obtain a \gls{rWERR} of \SI{5.8}{\percent} and \SI{6.3}{\percent} when using one past plus one future and two past contextual utterances, respectively.
It is particularly noteworthy that the improvements due to context are larger with a dual-mode approach than with vanilla \gls{RNN-T} training for streaming models, thus, justifying the proposed method.

\cref{table:dual_mode_context_training_2p_0f_raelr} shows \glspl{rAELR} relative to a streaming model without context (first line).
The proposed dual-mode contextual-utterance training is able to achieve a \gls{rAELR} of more than \SI{40}{ms}.
This is due to the use of the in-place lattice distillation loss \cite[Table 4, Data Row 2]{distillation} which helps the streaming system to imitate the behaviour of the teacher mode, and hence, reducing the emission latency of the last token with respect to the classical \gls{RNN-T} training.
The previous tuning results in \cref{table:beta} further confirm this hypothesis.

\section{Conclusions}
\label{sec:conclusions}

In this work we employed two techniques for training \gls{RNN-T}-based \gls{ASR} systems: (i) contextual-utterance training which makes use of the available contextual utterances; and (ii) dual-mode contextual-utterance training which allows to distill knowledge from a non-streaming mode with access to all utterances into a streaming mode with access to past utterances up to the current frame.
Using a state-of-the-art conformer transducer model, the proposed training techniques lead to a \gls{WER} reduction of \SI{12.1}{\percent} and \SI{6.3}{\percent} relative over standard \gls{RNN-T} training without context for non-streaming and streaming systems, respectively. Moreover, the dual-mode contextual-utterance training technique allows to maintain a relative last token emission latency reduction of over \SI{40}{ms}.
\clearpage
\balance
\bibliographystyle{IEEEtran}

\end{document}